\documentclass{sigchi-ext}
\usepackage[T1]{fontenc}
\usepackage{textcomp}
\usepackage[scaled=.92]{helvet} 
\usepackage{graphicx} 
\usepackage{balance}  
\usepackage{booktabs} 
\usepackage{ccicons}  
\usepackage{ragged2e} 
\usepackage{dirtytalk}

\def\plaintitle{Context, Utility and Influence of an Explanation} 
\def\emptyauthor{}
\def\plainkeywords{
  Explainable AI; Decision Theory; Contextual Utility Theory}

\title{Context, Utility and Influence of an Explanation}

\numberofauthors{6}
\author{%
  \alignauthor{%
    \textbf{Minal Suresh Patil}\\
    \affaddr{Umeå universitet} \\
    \email{minalsp@cs.umu.se} }\alignauthor{%
    \textbf{Kary Främling}\\
    \affaddr{Umeå universitet}\\
    \email{kary.framling@cs.umu.se} } \vfil \alignauthor{%
    } }

\definecolor{linkColor}{RGB}{6,125,233}
\hypersetup{%
  pdftitle={\plaintitle},
  pdfauthor={\emptyauthor},
  pdfkeywords={\plainkeywords},
  bookmarksnumbered,
  pdfstartview={FitH},
  colorlinks,
  citecolor=black,
  filecolor=black,
  linkcolor=black,
  urlcolor=linkColor,
  breaklinks=true,
}

\begin{document}

\CopyrightYear{2020}
\setcopyright{rightsretained}
\conferenceinfo{CHI'20,}{April  25--30, 2020, Honolulu, HI, USA}
\isbn{978-1-4503-6819-3/20/04}
\doi{https://doi.org/10.1145/3334480.XXXXXXX}
\copyrightinfo{\acmcopyright}

\maketitle

\RaggedRight{} 

\begin{abstract}
Contextual utility theory integrates context-sensitive factors into utility-based decision-making models. It stresses the importance of understanding individual decision-makers' preferences, values, and beliefs and the situational factors that affect them. Contextual utility theory benefits explainable AI. First, it can improve transparency and understanding of how AI systems affect decision-making. It can reveal AI model biases and limitations by considering personal preferences and context. Second, contextual utility theory can make AI systems more personalized and adaptable to users and stakeholders. AI systems can better meet user needs and values by incorporating demographic and cultural data. Finally, contextual utility theory promotes ethical AI development and social responsibility. AI developers can create ethical systems that benefit society by considering contextual factors like societal norms and values. This work, demonstrates how contextual utility theory can improve AI system transparency, personalization, and ethics, benefiting both users and developers.
\end{abstract}




\section{Introduction}
Decision theory and utility theory are important concepts in the field of explainable artificial intelligence (XAI)\cite{framling2020decision,wang2019designing,framling2022contextual}. These theories provide a framework for making informed and rational decisions in uncertain environments.

Decision theory is a branch of mathematics that deals with the process of making decisions in situations where there is uncertainty about the outcomes of different choices\cite{fishburn1968utility, dyer1992multiple, wallenius2008multiple, von1975multi}. It provides a systematic approach for analyzing and choosing between different options based on the available information and the decision-maker's preferences. Decision theory considers factors such as risk, uncertainty, and the potential outcomes of each choice in order to determine the best course of action.

Utility theory is a sub field of decision theory that focuses on the quantification of preferences\cite{stigler1950development,figueira2005maut,dyer2016multiattribute,bell1986or}. It assumes that individuals have a preference ordering over different outcomes, and that these preferences can be represented as a numerical value, known as utility. The utility of an outcome is calculated based on the individual's preferences and the decision-maker's assessment of the likelihood of different outcomes. By combining decision theory and utility theory, organisations can make informed and rational decisions that reflect their goals and values\cite{haring1959utility,hagen1995risk,sebora1995expected}.

To understand the concept of utility, it is helpful to consider an example. Suppose you are trying to decide between two job offers, one that pays more but requires a longer commute and one that pays less but is closer to your home. The decision involves two attributes, salary and commute time, and you must decide which attribute is more important to you.
To evaluate the job offers, you must first assign utility scores to each attribute at each level. For example, you might assign a utility score of $0.8$ to a salary of $\$80,000$ and a utility score of $0.5$ to a salary of $\$60,000$. Similarly, you might assign a utility score of $0.5$ to a commute time of $30$ minutes and a utility score of $0.3$ to a commute time of $60$ minutes.

Once you have assigned utility scores to each attribute at each level, you can calculate the overall utility of each job offer. This involves weighting the utility scores for each attribute by their relative importance, which is typically expressed as a percentage. For example, you might assign a weight of $60\%$ to salary and a weight of $40\%$ to commute time, reflecting your priorities.

The overall utility of each job offer is then calculated as a weighted sum of the utility scores for each attribute, using the weights assigned to each attribute. For example, the overall utility of the first job offer might be calculated as follows:

\begin{equation}
    0.8 * 0.6 + 0.5 * 0.4 = 0.68
\end{equation}

This means that the first job offer has an overall utility of $0.68$, which reflects how well it satisfies your preferences and goals.

The process of assigning utility scores and weights to attributes is subjective, and different decision-makers may assign different scores and weights based on their preferences and goals. This subjectivity is one of the strengths of utility theory, as it allows decision-makers to express their priorities in a meaningful way.

In the context of XAI, decision theory and utility theory play an important role in ensuring that AI systems are transparent and explainable\cite{framling122021contextual}. By incorporating these theories into AI systems, organisations can ensure that decisions are made in a systematic and consistent manner, and that the reasoning behind decisions is easily understood. This helps to build trust in the AI system and ensures that decisions are aligned with the goals and values of the organisation.



\section{Importance and Influence}
In utility theory, \say{importance} and \say{influence} are both key concepts that are used to make decisions based on multiple criteria.

\textbf{Importance} refers to the degree to which each attribute is valued or weighted in the decision-making process. In other words, it represents the decision maker's preferences or priorities for each attribute. Importance is typically expressed as a numerical value or a weight, and it can be subjective or objective depending on the decision maker's goals and objectives.

For example, in choosing a new car, one person may consider fuel efficiency to be the most important attribute, and so they might assign a higher weight to it in the decision-making process. Another person may prioritize safety, and so they would assign a higher weight to the car's safety rating.

\textbf{Influence} refers to the degree to which each attribute affects the overall utility or value of the decision. In other words, it represents the objective or subjective impact of each attribute on the decision. Influence can be expressed as a number or a ranking, and it is usually determined based on empirical data, expert opinions, or the decision maker's past experience.

For example, in the car-buying scenario, fuel efficiency might have a greater influence on the overall value of the car than the safety rating or the price. This would mean that changes in the fuel efficiency would have a bigger impact on the decision than changes in the safety rating or price. 

In summary, importance and influence are both used in utility theory to help decision-makers make choices based on multiple criteria. Importance reflects the decision maker's priorities, while influence reflects the objective or subjective impact of each attribute on the decision.

\section{Contextual Utility Theory in XAI}
Contextual Utility Theory (CUT) is a framework that allows for the development of XAI models. CUT focuses on the context in which decisions are made, and how this context can be used to provide explanations for the output of an AI model. In this work, we will explore the principles of CUT, how it can be used to develop XAI models, and provide examples of its application in various domains.

\subsection{Principles of Contextual Utility Theory}
The fundamental principle of CUT is that the utility of a decision is not fixed, but rather a function of the context in which the decision is made. This means that the same decision can have different utilities in different contexts. For example, the decision to purchase a car may have a different utility for a single person living in the city compared to a family living in a rural area. This is because the context in which the decision is being made affects the value that the decision provides. By integrating user-specific information, such as demographic or cultural factors, AI systems can tailor their outputs to better align with the needs and values of their users. Finally, contextual utility theory can help promote ethical decision-making and social responsibility in AI development. By accounting for contextual factors such as societal norms or values, AI developers can design systems that align with broader ethical principles and promote positive social outcomes and this has been further elaborated in this position paper from the previous year's edition \cite{singh2022grounding} which emphasis that explainability should be grounded in social contexts.

CUT provides a way to account for context in decision-making by using contextual variables. Contextual variables are variables that describe the context in which a decision is made. For example, in the loan approval example, contextual variables may include the credit score of the applicant, the applicant's employment status, and the purpose of the loan. These variables are used to calculate the utility of a decision, which is then used to determine the best course of action.
The CUT framework also incorporates the principle of explainability. Explainability refers to the ability of an AI model to provide a clear and understandable explanation for its decisions. By using contextual variables to calculate utility, CUT allows for the creation of models that can provide explanations that are tied to the specific context in which the decision is being made. This makes it easier for users to understand the decision-making process of the model and to trust its output.  
CUT can be mathematically represented using a utility function that takes into account the individual's preferences and the context in which the decision is being made. The utility function:
\begin{equation}
    U(x,c) = u(x) + v(c)
\end{equation}

where $U$ is the overall utility, x is the decision variable, $c$ is the contextual variable, $u(x)$ is the individual's preference function for the decision variable, and $v(c)$ is the contextual value function.

For example, consider a restaurant recommendation AI system that takes into account the user's preferences and the contextual factors such as the time of day and the user's location. The utility function for this AI system could be represented as:

\begin{equation}
    U(x,c) = u(x) + v(c) = \alpha1(r) + \alpha2(p) + \beta1(t) + \beta2(d)
\end{equation}

where $x$ is the recommendation (i.e., a particular restaurant), $c$ is the context, $r$ is the restaurant's rating, $p$ is the price, $t$ is the time of day, and $d$ is the distance to the restaurant. The parameters $\alpha$ and $\beta$ represent the weights assigned to each factor based on the user's preferences.

The user's preference function $u(x)$ can be represented as:

\begin{equation}
    u(x) = \alpha1(r) + \alpha2(p)
\end{equation}

where $\alpha1$ and $\alpha2$ are the weights assigned to the restaurant's rating and price, respectively. The contextual value function $v(c)$ can be represented as:

\begin{equation}
    v(c) = \beta1(t) + \beta2(d)
\end{equation}

where $\beta1$ and $\beta2$ are the weights assigned to the time of day and distance to the restaurant, respectively.

Using this utility function, the AI system can recommend restaurants that have the highest utility value based on the user's individual preferences and the specific context in which the decision is being made. By providing an interpretable and transparent decision-making process, this approach can help users understand why certain recommendations are being made and build trust in the AI system.

\subsection{Using Contextual Utility Theory in Explainable AI}
CUT can be used to develop XAI models in a variety of domains, including finance, healthcare, and autonomous vehicles. In each domain, contextual variables can be used to calculate the utility of a decision and provide explanations for the model's output.

\begin{itemize}
    \item \textbf{Loan Approval} - One example of the application of CUT in finance is the loan approval process. In the loan approval process, a lender must decide whether to approve or reject a loan application. To make this decision, the lender considers various factors such as the applicant's credit score, employment status, and the purpose of the loan. To develop an XAI model using CUT, the lender could use these factors as contextual variables. The model could then use these variables to calculate the utility of approving or rejecting the loan. The output of the model could then be explained based on the context of the application. For example, the model may explain that a loan was rejected because the applicant's credit score was too low or because the purpose of the loan did not align with the lender's policies.
    \item \textbf{Medical Diagnosis} - Another example of the application of CUT is in the field of healthcare. In healthcare, a physician must make a diagnosis based on the patient's symptoms and medical history. The physician may use a variety of diagnostic tests to help with this process.
To develop an XAI model using CUT, the physician could use the patient's medical history, age, and other relevant factors as contextual variables. The model could then use these variables to calculate the utility of different diagnoses. The output of the model could then be explained based on the context of the patient. For example, the model may explain that a particular diagnosis was made because the patient had a history of similar symptoms or because the patient was of a certain age.
    
\end{itemize}

\section{Conclusion}
In conclusion, contextual utility theory offers a valuable framework for developing explainable AI systems. By taking into account the various contextual factors that influence decision-making, such as user preferences and the nature of the task at hand, this theory provides a more nuanced approach to modeling and interpreting AI outputs.

One of the key benefits of contextual utility theory is that it enables AI systems to generate more personalized recommendations and decisions. By incorporating individual preferences and situational factors, these systems can better adapt to the needs of different users and deliver more tailored outputs. Moreover, contextual utility theory can help to identify potential biases and limitations in AI systems, and enable developers to mitigate these issues in order to improve overall performance and reliability.

Overall, contextual utility theory represents a promising direction for the development of explainable AI systems. As the field continues to evolve, it will be important to explore further how this theory can be applied in practice and refined to improve the accuracy, transparency, and ethical implications of AI decision-making. By embracing these principles, we can work towards creating more trustworthy and effective AI systems that benefit society as a whole.


\section*{Acknowledgements}
The author thanks Timotheus Kampik for guidance and valuable insights in this work. This work was partially funded by the Knut and Alice Wallenberg Foundation.
\balance{} 



\end{document}